\documentclass[showpacs,twocolumn,aps,prl,superscriptaddress,letterpaper]{revtex4-1}

\pdfoutput=1

\usepackage{graphicx}
\usepackage[caption=false]{subfig}
\usepackage{amsmath}
\usepackage{color}

\newcommand{\rev}[1]{\textcolor{black}{#1}}

\usepackage[utf8]{inputenc}
\definecolor{mpip} {rgb}{0,0.490196078431,0.478431372549}
\definecolor{mpip2} {rgb}{0,0.690196078431,0.278431372549}
\definecolor{grau} {rgb}{.5,.5,.5}
\definecolor{refokay} {rgb}{.1,.9,.7}

\begin{document}


\title{
\rev{Forces on} rigid inclusions in elastic media \rev{and resulting} matrix-mediated interactions
}

\author{Mate Puljiz}
\email{puljiz@thphy.uni-duesseldorf.de}
\affiliation{Institut f{\"u}r Theoretische Physik II: Weiche Materie, 
Heinrich-Heine-Universit{\"a}t D{\"u}sseldorf, D-40225 D{\"u}sseldorf, Germany}
\author{Shilin Huang}
\email{huangs@mpip-mainz.mpg.de}
\affiliation{Max Planck Institute for Polymer Research, Ackermannweg 10, 55128 Mainz, Germany}
\author{G\"unter K. Auernhammer}
\email{auhammer@mpip-mainz.mpg.de}
\affiliation{Max Planck Institute for Polymer Research, Ackermannweg 10, 55128 Mainz, Germany}
\author{Andreas M. Menzel}
\email{menzel@thphy.uni-duesseldorf.de}
\affiliation{Institut f{\"u}r Theoretische Physik II: Weiche Materie, 
Heinrich-Heine-Universit{\"a}t D{\"u}sseldorf, D-40225 D{\"u}sseldorf, Germany}

\date{see https://doi.org/10.1103/PhysRevLett.117.238003}



\begin{abstract}
To describe many-particle systems suspended in incompressible low-Reynolds-number \textit{fluids}, effective hydrodynamic interactions can be introduced. Here, we consider particles embedded in 
\textit{elastic} media. The effective elastic 
interactions between spherical particles are calculated analytically, 
inspired by the approach in the fluid case.
Our experiments on interacting magnetic particles confirm the theory.
In view of the huge success of the method in hydrodynamics, we similarly expect many future applications in the elastic case, e.g.\ for elastic composite materials. 
\end{abstract}

\maketitle


Hydrodynamics determines our daily life. 
Examples are given by the flow of air into our lungs \cite{zhang2002transient}, drinking of beverages and digestive processes \cite{meng2005computer,*ferrua2010modeling}, technical applications such as microfluidic devices \cite{squires2005microfluidics}, or shape optimization of planes, vehicles, ships, and propellers
\cite{wald2006aerodynamics,*campana2006shape,*muller2014aerodynamic}. 
All these processes are described by the 
Navier-Stokes equations \cite{navier1822memoire,*stokes1845theories} or variants thereof. 
This set of equations typically poses significant challenges during solution due to a convective nonlinearity reflecting inertial 
effects. 
Basically, turbulence is driven by the inertial term. 
It often renders analytical solutions impossible. 

The situation changes for small dimensions and velocities or high viscosity. 
Then, the relative strength of inertial effects, measured by the Reynolds number, is low. The nonlinearity can be neglected. 
A Green's function in terms of the so-called Oseen matrix is then available, which formally solves the  
problem analytically \cite{karrila1991microhydrodynamics,dhont1996introduction}.
In this way, semi-dilute colloidal suspensions, i.e.\ the dispersion of nano- to micrometer-sized particles in a fluid \cite{dhont1996introduction,felderhof1977hydrodynamic,*ermak1978brownian,*durlofsky1987dynamic, *zahn1997hydrodynamic,*meiners1999direct,*dhont2004thermodiffusion,*rex2008influence}, or microswimmer suspensions 
\cite{pooley2007hydrodynamic,*baskaran2009statistical,*menzel2016dynamical, *lauga2009hydrodynamics,*drescher2010direct,*drescher2011fluid,*paxton2004catalytic} are described effectively. The explicit role of the fluid is eliminated and replaced by effective hydrodynamic interactions between the suspended particles 
\cite{karrila1991microhydrodynamics,dhont1996introduction}. 

Despite the success of this theoretical approach for colloidal suspensions, hardly any investigations consider a surrounding elastic solid instead of a suspending fluid. 
This is surprising, since, as we show below, the formalism can be adapted straightforwardly to linearly elastic matrices and is confirmed by our experiments. 
Our approach will, for instance, facilitate describing the response of elastic composite materials to external stimuli. Such materials consist of more or less rigid inclusions embedded in an elastic matrix. They are of growing technological interest and may serve, e.g., as soft actuators or sound attenuation devices \cite{an2003actuating,*fuhrer2009crosslinking,*bose2012soft, *cheng2006observation,*still2011collective, *baumgartl2007tailoring,*baumgartl2007erratum}.

In previous theoretical studies, the physics of one single rigid or deformable inclusion was addressed \cite{eshelby1957determination,*eshelby1959elastic, *walpole1991rotated,*walpole1991translated,*walpole2005green,phanthien1993rigid}, also under acoustic irradiation \cite{oestreicher1951field,*norris2006impedance,*norris2008faxen}. 
For more than a single inclusion, mainly the so-called load problem was analyzed theoretically for a pair of rigid inclusions: one prescribes displacements of two rigid inclusions in an elastic matrix, and then determines the forces necessary to achieve these given displacements \cite{phanthien1994loadtransfer,*kim1995faxen}. 

Here, we take the converse point of view,
based on the cause-and-effect chain in our 
experiments: 
external forces are imposed onto the inclusions, or mutual forces between the inclusions are induced, for example to 
actuate the material or to tune its properties. 
In response to the forces, the inclusions are displaced. Since they cannot penetrate through the surrounding elastic matrix,
they transmit the forces to the matrix and distort it. 
Such distortions lead to mutual long-ranged interactions between the inclusions, in analogy to hydrodynamic interactions in colloidal suspensions \cite{karrila1991microhydrodynamics,dhont1996introduction,tanaka2000simulation}. 

We present a basic derivation of 
analytical expressions for these interactions from the underlying elasticity equations. Then, we verify the theory by 
experiments 
on rigid paramagnetic particles embedded in soft elastic matrices. 
Mutual particle interactions 
are induced by an external magnetic field. As we demonstrate, 
theory and experiment are in good agreement, \rev{and also allow for microrheological measurements \cite{ziemann1994local,*bausch1999measurement,
*waigh2005microrheology,*wilhelm2008out}}.

For simplicity, we assume a homogeneous, isotropic, infinitely extended elastic matrix, and low-\-am\-pli\-tude
deformations. 
Applying a bulk force density $\mathbf{f}_b(\mathbf{r})$ to the matrix, its equilibrated state satisfies the linear elastostatic Navier-Cauchy equations \cite{cauchy1828exercises}, 
\begin{equation}\label{eq_navier-cauchy}
	\nabla^2\mathbf{u}(\mathbf{r}) + \frac{1}{1-2\nu}\nabla\nabla\cdot\mathbf{u}(\mathbf{r}) ={} -\frac{1}{\mu}\mathbf{f}_b(\mathbf{r}).
\end{equation}
This is the elastic analogue to the linearized Stokes equation in low-Reynolds-number hydrodynamics \cite{dhont1996introduction}. 
Instead of velocities, $\mathbf{u}(\mathbf{r})$ here denotes the displacement field, describing the reversible relocations of the volume elements from their initial positions during deformations. 
$\mu$ is the shear modulus of the matrix and $\nu$ its Poisson ratio, connected to its compressibility \cite{landau1986theory}. 
\rev{We consider an incompressible matrix, i.e.\ $\nabla\cdot\mathbf{u}(\mathbf{r})=0$ along with $\nu=0.5$. Yet, in contrast to the hydrodynamic case, also compressible elastic systems are readily addressed, and we present the corresponding expressions in the supplemental material \cite{supplinfo}.} 

Importantly, for a point force density $\mathbf{f}_b(\mathbf{r})=\mathbf{F}\delta(\mathbf{r})$ acting on the matrix, the resulting deformation field can be calculated analytically from Eq.~(\ref{eq_navier-cauchy}) via Fourier transform as $\mathbf{u}(\mathbf{r})=\mathbf{\hspace{.02cm}\underline{\hspace{-.02cm}G}}(\mathbf{r})\cdot\mathbf{F}$. Here,
\begin{equation}\label{greens_function}
	\mathbf{\hspace{.02cm}\underline{\hspace{-.02cm}G}}(\mathbf{r}) ={} \frac{1}{\rev{8\pi}\mu}\left[\frac{\rev{1}}{r}\mathbf{\underline{\hat{I}}}+\frac{\mathbf{r}\mathbf{r}}{r^3}\right]
\end{equation}
is the corresponding Green's function \cite{landau1986theory}, $\mathbf{\underline{\hat{I}}}$ the identity matrix, $r$=$|\mathbf{r}|$, and the underscore marks second-rank tensors and matrices. 
Still, it is practically impossible to explicitly solve Eq.~(\ref{eq_navier-cauchy}) analytically in the presence of several rigid embedded particles of finite size. 
An iterative procedure resolves this problem, \rev{see Fig.~\ref{fig_illustration}}. 
\begin{figure}
\centerline{\includegraphics[width=6.5cm]{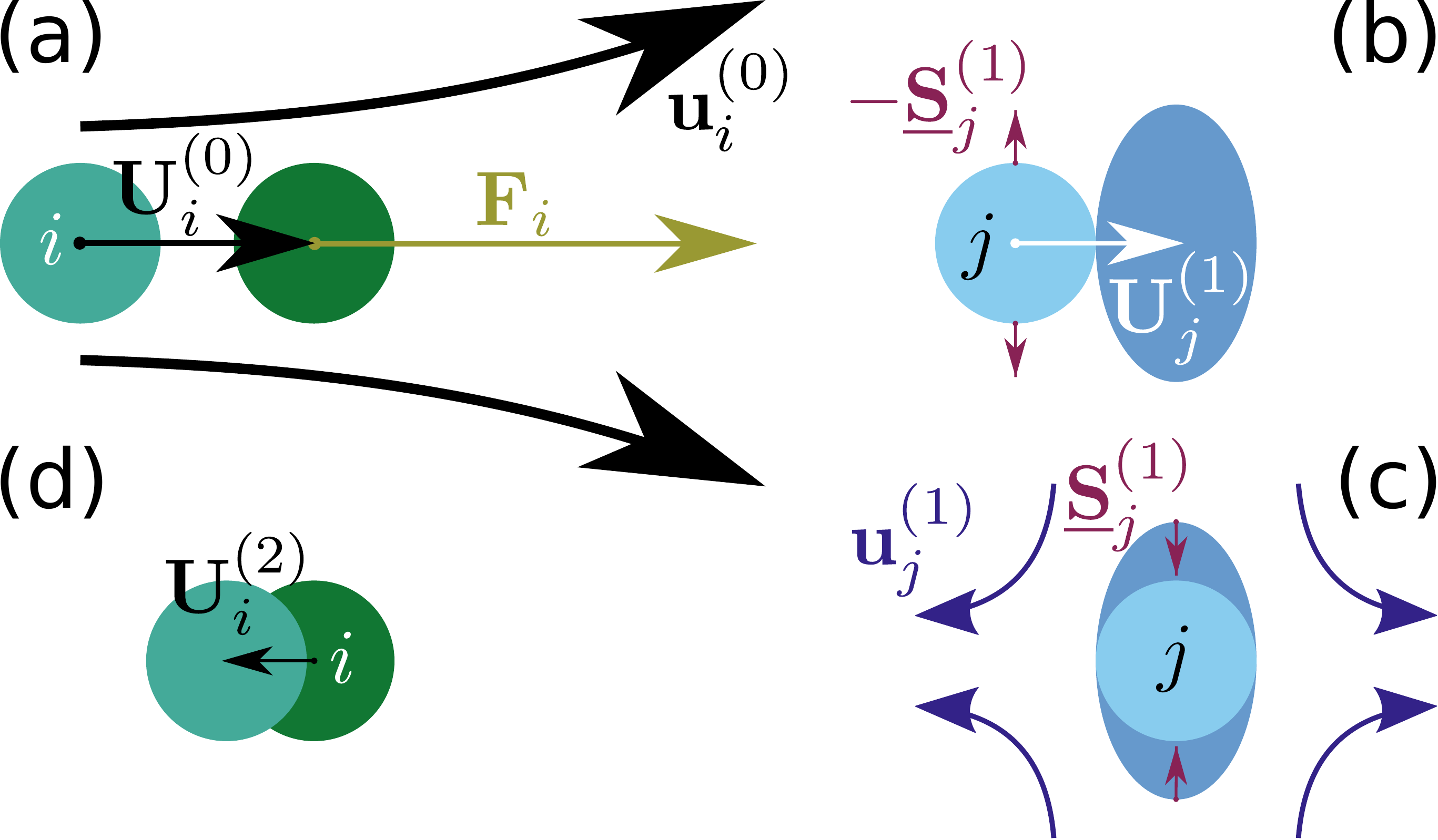}}
\caption{
\rev{Illustration of the reflection of a displacement field $\mathbf{u}_i^{(0)}$ induced by (a) the force $\mathbf{F}_i$ that displaces particle $i$ by $\mathbf{U}_i^{(0)}$. (b) Due to $\mathbf{u}_i^{(0)}$, particle $j$ gets displaced by $\mathbf{U}_j^{(1)}$ and would be distorted as described by a stresslet $-\mathbf{\underline{S}}_j^{(1)}$ (rotations $\mathbf{\Omega}_j^{(1)}$ not depicted for simplicity). (c) However, particle $j$ is rigid and resists to deformation, expressed by a counteracting stresslet $\mathbf{\underline{S}}_j^{(1)}$, which results in a displacement field $\mathbf{u}_j^{(1)}$. (d)~The reflected $\mathbf{u}_i^{(0)}$, i.e.\ $\mathbf{u}_j^{(1)}$, 
displaces particle $i$ by $\mathbf{U}_i^{(2)}$.}
}
\label{fig_illustration}
\end{figure}

We consider $N$ rigid spherical particles of radius $a$, with no-slip boundary conditions on their surfaces. 
First we only address the $i$th particle 
at position $\mathbf{r}_i$, subject to an external force $\mathbf{F}_i$. 
The embedded particle transmits this force to the surrounding matrix and induces a displacement field 
\begin{equation}\label{eq_u_i_0}
	\mathbf{u}_i^{(0)} (\mathbf{r}) = \left(1+\frac{a^2}{6}\nabla^2\right)\mathbf{\hspace{.02cm}\underline{\hspace{-.02cm}G}}(\mathbf{r}-\mathbf{r}_i)\cdot\mathbf{F}_i.
\end{equation}
This field is the elastic analogue of hydrodynamic Stokes flow 
\cite{karrila1991microhydrodynamics,dhont1996introduction}, 
for 
elastic media.  
Inserting Eq.~(\ref{greens_function}) reproduces a corresponding expression in Ref.~\onlinecite{phanthien1993rigid}. 
Eq.~(\ref{eq_u_i_0}) is confirmed 
as it satisfies Eq.~(\ref{eq_navier-cauchy}), shows the correct limit $\mathbf{u}_i^{(0)}(\mathbf{r})=\mathbf{\hspace{.02cm}\underline{\hspace{-.02cm}G}}(\mathbf{r}\rev{-\mathbf{r}_i})\cdot\mathbf{F}_i$ for \rev{$|\mathbf{r}-\mathbf{r}_i|>a$ when} $a\rightarrow0$, and for $\rev{|\mathbf{r}-\mathbf{r}_i|}=a$ is constant on the particle surface. 
Thus, Eq.~(\ref{eq_u_i_0}) for $\rev{|\mathbf{r}-\mathbf{r}_i|}=a$ reveals the rigid displacement 
\begin{equation}
\mathbf{U}_i^{(0)} \,=\, \mathbf{u}_i^{(0)}(\rev{|\mathbf{r}-\mathbf{r}_i|}=a) \,=\, \frac{\rev{1}}{\rev{6\pi}\mu a}\mathbf{F}_i
\end{equation} 
of the $i$th particle in response to $\mathbf{F}_i$ in accord with the no-slip 
conditions at $\rev{|\mathbf{r}-\mathbf{r}_i|}=a$. 

To find the effective elastic interactions between particles $i$ and $j$ ($j\neq i$), we take the induced displacement field $\mathbf{u}_i^{(0)}(\mathbf{r})$ as given. 
We need to determine how particle $j$ reacts to the imposed field $\mathbf{u}_i^{(0)}(\mathbf{r})$. 
In general, particle $j$ can be rigidly translated by a displacement vector $\mathbf{U}_j^{(1)}$ and rigidly rotated by a rotation vector $\mathbf{\Omega}_j^{(1)}$. 
Taking into account the no-slip 
conditions on the surface $\partial V_j$ of the $j$th particle, the equality 
\begin{equation}\label{eq_balance}
\mathbf{U}_j^{(1)}+\mathbf{\Omega}_j^{(1)}\times(\mathbf{r}-\mathbf{r}_j)=\mathbf{u}_i^{(0)}(\mathbf{r})+\int_{\partial V_j}\mathbf{\hspace{.02cm}\underline{\hspace{-.02cm}G}}(\mathbf{r}-\mathbf{r}')\cdot\mathbf{f}(\mathbf{r}')\mathrm{d}S'
\end{equation}
must hold for all $\mathbf{r}\in\partial V_j$. \rev{That is, the rigid displacement of each point on the surface shell of particle $j$ 
(l.h.s.) must equal the displacement field in the matrix at the same point (r.h.s.). 
The latter is given by the imposed displacement field, here $\mathbf{u}_i^{(0)}(\mathbf{r})$, plus the deformation that the particle surface itself induces in the matrix, i.e.\ the integral term.}
\rev{Also an externally imposed global displacement field could be included (on the r.h.s.).}
$\mathbf{f}(\mathbf{r}')$ 
describes the surface force density exerted by the surface 
of particle $j$ onto the matrix.

Such an embedded particle will translate and rotate as dictated by the surrounding matrix. 
We obtain the expression for $\mathbf{U}_j^{(1)}$ by integrating Eq.~(\ref{eq_balance}) over $\partial V_j$. 
Similarly, for $\mathbf{\Omega}_j^{(1)}$, Eq.~(\ref{eq_balance}) is 
multiplied dyadically by $\mathbf{r}-\mathbf{r}_j$, and after integration over $\partial V_j$ the antisymmetric part is extracted. 
\rev{To perform the calculation}, 
$\mathbf{u}_i^{(0)}(\mathbf{r})$ is Taylor expanded around $\mathbf{r}_j$. 
\rev{Moreover}, we use that Eq.~(\ref{eq_navier-cauchy}) for $\mathbf{r}\notin\partial V_i$ leads to \rev{$\nabla^4\mathbf{u}_i^{(0)}(\mathbf{r})=\mathbf{0}$ and $\nabla\times\nabla^2\mathbf{u}_i^{(0)}(\mathbf{r})=\mathbf{0}$}. 
The last term in Eq.~(\ref{eq_balance}) vanishes at this stage as no total net external force or torque is applied to particle $j$ at the present step of iteration. 
In the end, we recover the elastic analogues of the hydrodynamic \cite{batchelor1972hydrodynamic,karrila1991microhydrodynamics,dhont1996introduction} Fax\'en laws 
\begin{eqnarray}
		\mathbf{U}_j^{(1)} &={} &\left(1+\frac{a^2}{6}\nabla^2\right)\mathbf{u}_i^{(0)}(\mathbf{r})\bigg|_{\mathbf{r}=\mathbf{r}_j},\label{eq_faxen-translation} \\
		\boldsymbol{\Omega}_j^{(1)} &={} &\frac{1}{2}\nabla\times\mathbf{u}_i^{(0)}(\mathbf{r})\bigg|_{\mathbf{r}=\mathbf{r}_j}\label{eq_faxen-rotation}.
\end{eqnarray}
This is how particle $j$ is translated and rotated in the field $\mathbf{u}_i^{(0)}(\mathbf{r})$ induced by particle $i$. Yet, elastic retroaction occurs between the particles, as described in the following. 

The force densities $\mathbf{f}(\mathbf{r}')$ in Eq.~(\ref{eq_balance}) that the particles exert on their environment in general will not vanish identically. 
Since the particles are rigid, they resist \rev{any} deformation \rev{that $\mathbf{u}_i^{(0)}$ would imply}. 
Thus, they exert counteracting stresses onto the deformed matrix. 
The stresslet exerted by particle $j$ onto the matrix can be denoted as $\mathbf{\underline{S}}_j=\int_{\partial V_j}\mathrm{d}S'\{[\mathbf{f}(\mathbf{r}')\mathbf{r}'+
(\mathbf{f}(\mathbf{r}')\mathbf{r}')^T]/2 - \rev{\mathbf{\underline{\hat{I}}}\,[\mathbf{f}(\mathbf{r}')\cdot\mathbf{r}']/3} \}$, where $[\bullet]^T$ marks the transpose. 
In our case, we can directly calculate from Eq.~(\ref{eq_balance}) the stresslet $\mathbf{\underline{S}}_j^{(1)}$ that particle $j$ exerts onto the matrix when it resists to the deformation described by $\mathbf{u}_i^{(0)}(\mathbf{r})$. 
To find the expression for $\mathbf{\underline{{S}}}_j^{(1)}$, one proceeds in the same way as described above for $\mathbf{\Omega}_j^{(1)}$ but eventually extracts the symmetric part.  
The latter contains the definition of $\mathbf{\underline{S}}_j^{(1)}$. 
We obtain
\begin{equation}
	\mathbf{\underline{S}}_j^{(1)}\! = \frac{\rev{10\pi}\mu a^3}{-\rev{3}}\left(1+\frac{a^2}{10}\nabla^2\right)
\!
\left[ \nabla\mathbf{u}_i^{(0)}(\mathbf{r})+\big(\nabla\mathbf{u}_i^{(0)}(\mathbf{r})\big)^{\!T} \right]\!\Big|_{\mathbf{r}_j}\!\!.
\label{eq_faxen-stresslet}
\end{equation}
This stresslet leads to additional distortions of the matrix, \rev{see Fig.~\ref{fig_illustration}}, described by a displacement field $\mathbf{u}^{(1)}_j(\mathbf{r})$ that overlays $\mathbf{u}^{(0)}_i(\mathbf{r})$. 
We find $\mathbf{u}^{(1)}_j(\mathbf{r})$ from the general expression 
$\mathbf{u}_j(\mathbf{r})=\int_{\partial V_j}\mathrm{d}S'\mathbf{\hspace{.02cm}\underline{\hspace{-.02cm}G}}(\mathbf{r}-\mathbf{r}')\cdot\mathbf{f}(\mathbf{r}')$ by Taylor expanding the Green's function in $\mathbf{r}'$ around $\mathbf{r}'=\mathbf{r}_j$. 
The definition of $\mathbf{\underline{S}}_j$ shows up as the symmetric part of the second-order term of the series, similarly to the hydrodynamic case \cite{batchelor1972hydrodynamic,karrila1991microhydrodynamics}, leading to
\begin{equation}\label{eq_u_j_1}
	\mathbf{u}_j^{(1)}(\mathbf{r}) ={} -\left(\mathbf{\underline{S}}_j^{(1)}\cdot\nabla\right)\cdot\mathbf{\hspace{.02cm}\underline{\hspace{-.02cm}G}}(\mathbf{r}-\mathbf{r}_j).
\end{equation}

This expression completes our first step of iteration. 
In the second step, it is particle $i$ that is exposed to the field $\mathbf{u}_j^{(1)}(\mathbf{r})$. 
Correspondingly, we find its reaction from Eqs.~(\ref{eq_faxen-translation})--(\ref{eq_u_j_1}) by replacing $(\mathbf{u}_i^{(0)}, \mathbf{U}_j^{(1)}, \mathbf{\Omega}_j^{(1)}, \mathbf{\underline{S}}_j^{(1)}, \mathbf{u}_j^{(1)}, \mathbf{r}_j)$ with $(\mathbf{u}_j^{(1)}, \mathbf{U}_i^{(2)}, \mathbf{\Omega}_i^{(2)}, \mathbf{\underline{S}}_i^{(2)}, \mathbf{u}_i^{(2)}, \mathbf{r}_i)$. 
Particle $i$ now feels the consequences of its self-generated field $\mathbf{u}_i^{(0)}(\mathbf{r})$ \textit{reflected} by particle $j$ in the form of $\mathbf{u}_j^{(1)}(\mathbf{r})$. Therefore, the procedure was termed \textit{method of reflections} in hydrodynamics \cite{karrila1991microhydrodynamics,dhont1996introduction}. 
\rev{The displacement $\mathbf{U}_i^{(2)}$ 
in Fig.~\ref{fig_illustration} results from this reflection.} 
We have not found in the hydrodynamic derivation \cite{dhont1996introduction} the above reasoning of explicitly imposing on the matrix environment the rigidity-induced stress. 

In principle, this refinement of the deformation field via back-and-forth reflections between the two particles can be \rev{continued}, leading to increasingly-higher-order corrections in $a/r_{ij}$, where $r_{ij}=|\mathbf{r}_i-\mathbf{r}_j|$. 
For our example systems below, these iterations converge quickly, see Fig.~\ref{fig_two_particles}(c), so that it is sufficient to consider 
contributions up to (including) order $r_{ij}^{-4}$.

Due to the linearity of Eq.~(\ref{eq_navier-cauchy}), we can 
sum up the particle displacements obtained from the different steps of iteration. 
Moreover, we can consider external forces $\mathbf{F}_i$ on \textit{all} particles and 
calculate the resulting net displacements $\mathbf{U}_i$ due to the mutual elastic 
interactions 
($i=1,...,N$). 
These contributions 
superimpose. 
In analogy to the hydrodynamic \cite{dhont1996introduction} mobility matrix we 
express the result by an elastic \textit{displaceability matrix} $\mathbf{\underline{M}}$: 
%
%
%
\begin{eqnarray}\label{eq_displaceability_matrix}
\begin{pmatrix}
	\mathbf{U}_1\\
	\vdots\\
	\mathbf{U}_N
\end{pmatrix}
={}
\begin{pmatrix}
	\mathbf{\underline{M}}_{11} &   \ldots & \mathbf{\underline{M}}_{1N} \\
	\vdots & \vdots & \vdots  \\
	\mathbf{\underline{M}}_{N1} & \ldots & \mathbf{\underline{M}}_{NN}
\end{pmatrix}
\cdot
\begin{pmatrix}
	\mathbf{F}_1\\
	\vdots\\
	\mathbf{F}_N
\end{pmatrix}.
\end{eqnarray}
Limiting ourselves to contributions up to (including) order $r_{ij}^{-4}$, we find
\begin{eqnarray}
	\mathbf{\underline{M}}_{i=j} &\!=\!& M_0\Bigg[\mathbf{\underline{\hat{I}}} -  \sum\limits_{\substack{k=1 \\	k\not=i}}^{N} \frac{\rev{15}}{\rev{4}}\bigg(\frac{a}{r_{ik}}\bigg)^{\!\!\!4}  
\mathbf{\hat{r}}_{ik}\mathbf{\hat{r}}_{ik}
\Bigg], \label{M_II_2}\\
	\mathbf{\underline{M}}_{i\not=j} &\!=\!& M_0 \frac{3}{\rev{4}}\frac{a}{r_{ij}}  \Bigg[ \left( \rev{\mathbf{\underline{\hat{I}}}+} \mathbf{\hat{r}}_{ij}\mathbf{\hat{r}}_{ij} \right) 
   +  \rev{2} \bigg(\frac{a}{r_{ij}} \bigg)^{\!\!\!2} \!\left(\rev{\frac{1}{3}}\mathbf{\underline{\hat{I}}}-\mathbf{\hat{r}}_{ij}\mathbf{\hat{r}}_{ij}\right)\! \Bigg]
\notag\\ 
			& & {}+ \mathbf{\underline{M}}_{i\not=j}^{(3)}    ,
\label{M_IJ_2} 
\end{eqnarray}
where $M_0 ={} \rev{1}/\rev{6\pi}\mu a$ and $\mathbf{\hat{r}}_{ij}=\mathbf{r}_{ij}/r_{ij}$ ($i,j=1,2,...,N$). 

\rev{%
In Eq.~(\ref{M_II_2}), $\mathbf{\underline{\hat{I}}}$ represents the immediate displacement of particle $i$ due to the force $\mathbf{F}_i$ ($\mathbf{U}^{(0)}_i$ in Fig.~\ref{fig_illustration}). The second term $\sim\! r_{ik}^{-4}$ describes the rigidity-induced reflection from another particle ($\mathbf{U}^{(2)}_i$ in Fig.~\ref{fig_illustration}). It \textit{counteracts} $\mathbf{U}^{(0)}_i$, or vanishes for $\mathbf{F}_i\perp\mathbf{\hat{r}}_{ik}$. 
}

\rev{
In Eq.~(\ref{M_IJ_2}), $\rev{\mathbf{\underline{\hat{I}}}+} \mathbf{\hat{r}}_{ij}\mathbf{\hat{r}}_{ij}$ expresses the consequence of the force $\mathbf{F}_j$ acting on particle $j$: particle $i$ is relocated in the induced displacement field  
($\mathbf{U}^{(1)}_j$ in Fig.~\ref{fig_illustration}, for $i\leftrightarrow j$). The term $\sim\! r_{ij}^{-3}$ corrects this displacement field due to the finite size of particle $j$, in analogy to the Rodne-Prager formula in the fluid case \cite{dhont1996introduction}.
$\mathbf{\underline{M}}_{i\not=j}^{(3)}$ describes} additional three-body interactions $\sim\! r^{-4}_{ij}$ calculated in full analogy to the above procedure for the two-body interaction,
\begin{equation}\label{M_tt_3}
	\mathbf{\underline{M}}_{i\not=j}^{(3)} \!=\! M_0\frac{\rev{15}}{\rev{8}}
	\!\sum\limits_{\substack{k=1\\k\not=i,j}}^{N}
	\!\bigg(\frac{a}{r_{ik}}\bigg)^{\!\!\!2}
	\bigg(\frac{a}{r_{jk}}\bigg)^{\!\!\!2}
\Big[
\rev{1-3}(\mathbf{\hat{r}}_{ik}\cdot\mathbf{\hat{r}}_{jk})^2\Big]\mathbf{\hat{r}}_{ik}\mathbf{\hat{r}}_{jk}.
\end{equation}
That is, the deformation field induced by a force on a first particle \rev{$j$} spreads to a second particle \rev{$k$}, from where it is reflected towards the third particle \rev{$i$}. \rev{The angular dependence allows for configuration-dependent attractive, repulsive, or bypass contributions, see Fig.~\ref{fig_3body}.}
\begin{figure}
\centerline{\includegraphics[width=7.cm]{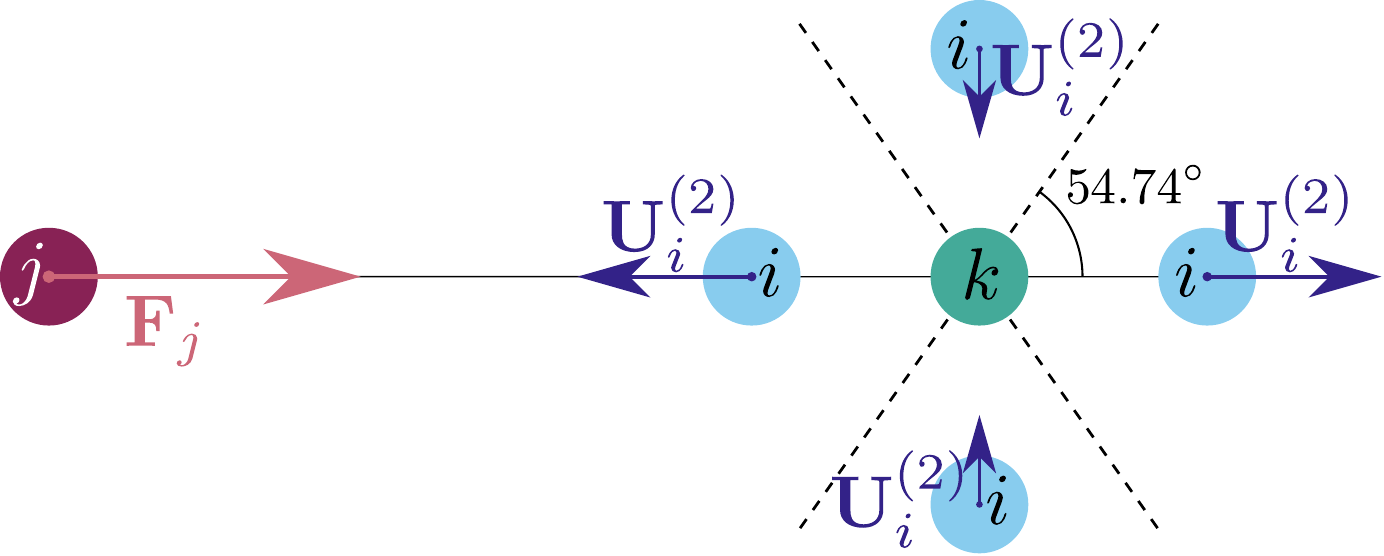}}
\caption{\rev{
Illustration of example three-body contributions in Eq.~(\ref{M_tt_3}). The force $\mathbf{F}_j$ on particle $j$ induces a displacement field that is reflected from particle $k$ due to its rigidity. Depending on its position, particle $i$ is effectively attracted or repelled by particle $j$ (strongest under coaxial alignment), pulled towards particle $k$ (bypass), or not affected at all (dashed). 
}}
\label{fig_3body}
\end{figure}

\begin{figure}
\centerline{\includegraphics[width=8.3cm]{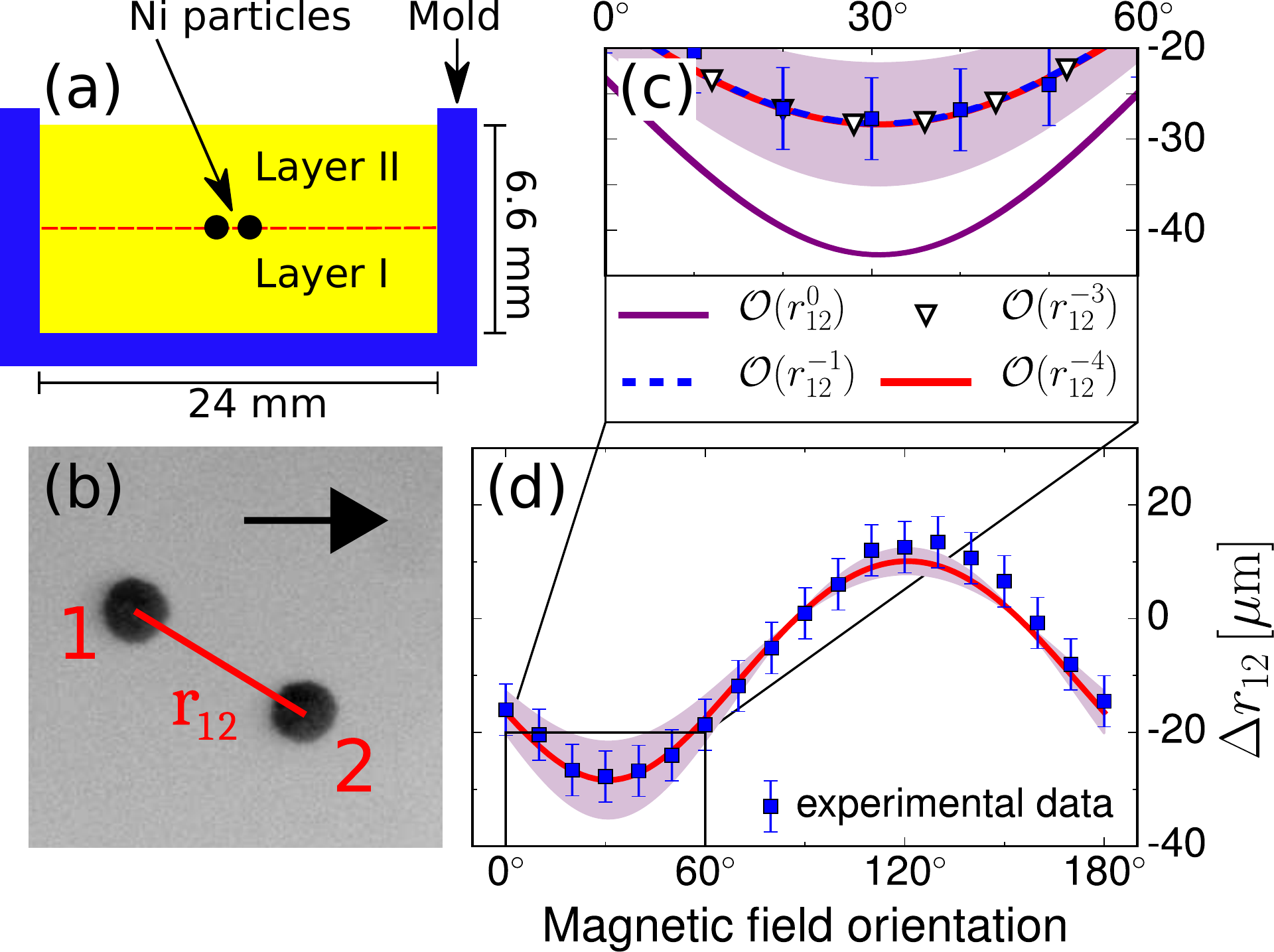}}
\caption{
(a) Schematic of the samples. After fabrication of the bottom gel layer (I), the paramagnetic nickel (Ni) particles are placed into the center plane (dashed), before the top layer (II) is added. The enclosing plastic molds are open to the top for optical investigation. 
(b) Snapshot of a system of two Ni particles (diameters 150.6$\pm$1.9~$\mathrm{\mu m}$) embedded in a soft elastic gel, here for vanishing external magnetic field. 
(c,d) Change in distance $\Delta r_{12}$ between the two particles
when applying an external magnetic field along different directions in the particle plane via clockwise rotation. The horizontal arrow in (b) defines the angle of~$0^{\circ}$.
Data points in (d) were measured experimentally. The line is calculated from the theory, where shaded areas arise from uncertainties in the experimental input values. 
\rev{An elastic modulus of $\mu={83.0}\pm{14.3}$~Pa is extracted.} 
The ``zoom'' in (c) highlights the rapid convergence of the theory. 
}
\label{fig_two_particles}
\end{figure}
Eqs.~(\ref{eq_displaceability_matrix})--(\ref{M_tt_3}) 
represent the central theoretical result. 
\rev{Up to (including) order $r_{ij}^{-4}$ it is exact, 
higher-body interactions for 
$N>3$ do not enter (see Ref.~\onlinecite{supplinfo} for $N=4$).}
To confirm and illustrate the merit of the 
theory, 
we performed experiments on small groups of paramagnetic particles embedded in a soft elastic gel matrix. 
Applying an external magnetic field induced mutual magnetic forces between the particles. Rotating the magnetic field 
tuned these forces. The resulting relative displacements of the particles were tracked by optical microscopy. 

We used paramagnetic Nickel (Ni) particles obtained from Alfa Aesar ($-100+325$ mesh, purity 99.8\%). The magnetic hysteresis curves 
(measured by a vibrating sample magnetometer, Lake Shore 7407) showed a low remanence of $\sim$ 7.5~kA/m,  a low coercive field of $\sim$~2.4~mT, and a volume magnetization of \rev{291}$\pm$17~kA/m under an external magnetic field of $\sim$~216~mT. We carefully selected Ni particles of similar sizes (deviation less than 2\% within each group) and a roundness 
$\gtrsim 0.91$ (measured by image analysis 
\cite{imagej}). These 
particles were embedded in the middle plane of a soft elastic polydimethylsiloxane-based \cite{huang2016buckling} gel, see Fig.~\ref{fig_two_particles}(a). 
First, a bottom gel layer 
with a thickness of 3.3~mm and a diameter of 24~mm was prepared in a plastic mold.  
Second, after 
sufficient stiffening
($\sim$~0.5~h), 
the Ni particles were carefully deposited on its top 
around the center. Third, a top gel layer 
with the same composition and size as 
the bottom layer was added. 
To ensure good connection between the two 
layers, 
at least 7 days of cross-linking were allowed. 

Using a 32-magnet Halbach array to generate a homogeneous magnetic field \cite{huang2016buckling}, we applied 
$\sim$ 216~mT to the embedded Ni particles, which is close to saturation. 
Starting from the initial direction, the magnetic field was rotated clockwise for $180^{\circ}$ in 18~steps within the 
plane containing the Ni particles. Their center-of-mass positions 
were tracked by a CCD camera (MATRIX VISION mvBlueCOUGAR-S) with the zoom macro lens (Navitar Zoom 7000) mounted above the samples and subsequent 
image analysis 
\cite{imagej}.

We measured the changes in particle distance $\Delta r_{ij}$ ($i\neq j$) for a two- and three-particle system, see Figs.~\ref{fig_two_particles} and \ref{fig_three_particles}, respectively, when rotating the external magnetic field. 
\begin{figure}
\centerline{\includegraphics[width=8.3cm]{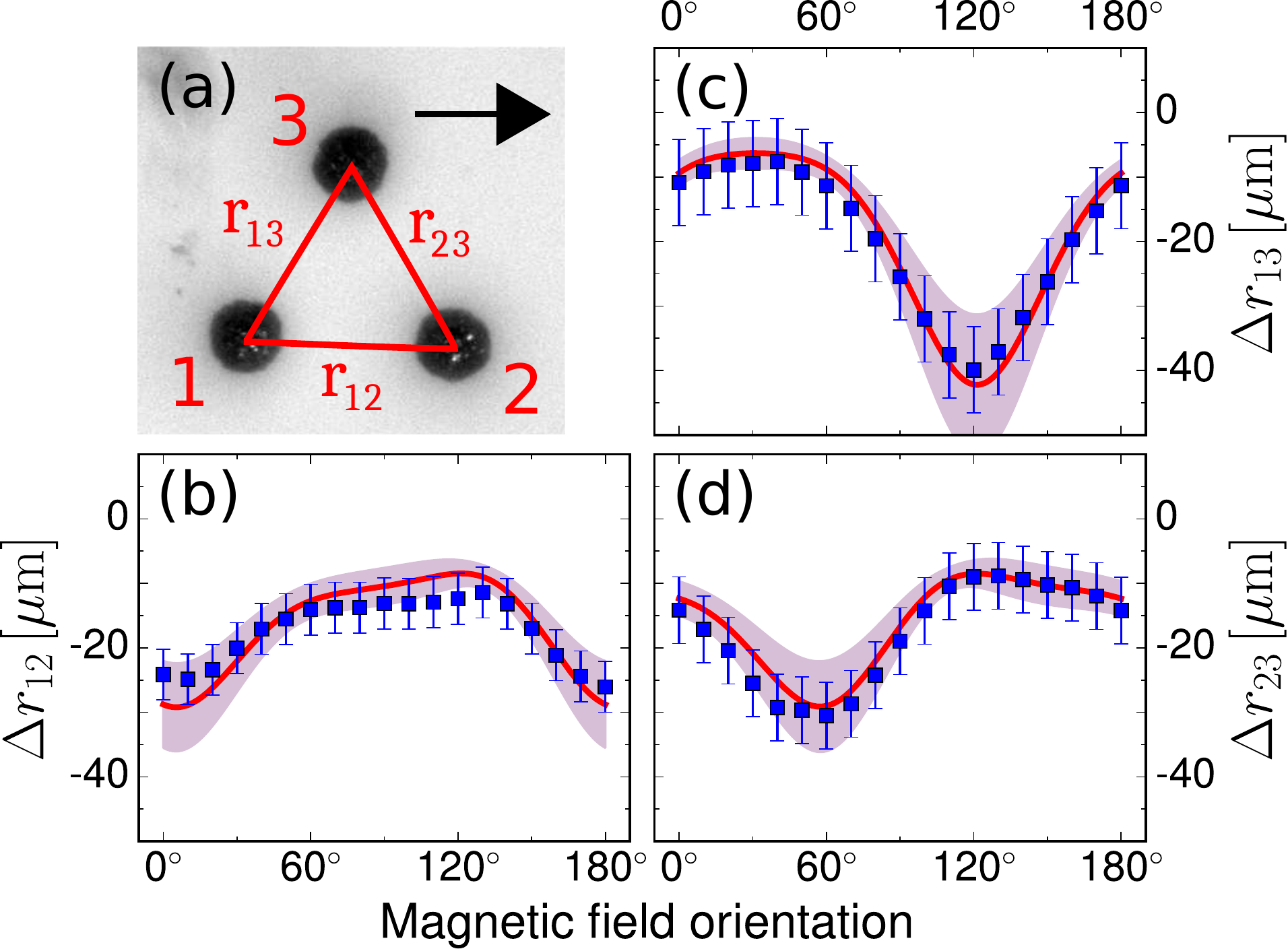}}
\caption{Same as in Fig.~\ref{fig_two_particles}(b) and (d), now for a three-particle system. (a) The snapshot was taken for vanishing external magnetic field (particle diameters 208.5$\pm$2.3~$\mathrm{\mu m}$). (b--d) Changes $\Delta r_{ij}$ in all three distances ($i,j=1,2,3$, $i\neq j$). \rev{The elastic modulus is $\mu={76.3}\pm{11.7}$~Pa.} } 
\label{fig_three_particles}
\end{figure}
Forces $\mathbf{F}_i$ on the particles 
result from mutual magnetic interactions. Due to substantial particle separations, we approximate the induced magnetic moments as point dipoles \cite{klapp2005dipolar,*biller2015mesoscopic}. 
Thus, we find 
\rev{\cite{jackson1962classical}}
\begin{equation}\label{eq_force}
	\mathbf{F}_i ={} -\frac{3\mu_0 m^2}{4\pi} \sum\limits_{\substack{j=1 \\ j\not=i}}^{N} \frac{ 5\mathbf{\hat{r}}_{ij}(\mathbf{\hat{m}}\cdot\mathbf{\hat{r}}_{ij})^2 - \mathbf{\hat{r}}_{ij}  - 2\mathbf{\hat{m}}(\mathbf{\hat{m}}\cdot\mathbf{\hat{r}}_{ij}) }{\rev{r_{ij}^{\:4}}} ,
\end{equation}
with $\mu_0$ the vacuum permeability and $\mathbf{m}=m\mathbf{\hat{m}}$ the induced magnetic moments, considered identical for all particles in the close-to-saturating homogeneous external magnetic field. 
%
%
Using as input parameters the experimentally determined particle positions, sizes, and magnetization, 
\rev{we extracted the elastic shear modulus} and calculated all changes $\Delta r_{ij}$ from Eqs.~(\ref{eq_displaceability_matrix})--(\ref{eq_force}). \rev{The magnetic forces $\mathbf{F}_i$ after displacement are determined iteratively.} Perfect agreement between theory and experiment in Figs.~\ref{fig_two_particles} and \ref{fig_three_particles} 
supports the significance of the theoretical approach \rev{and highlights its potential for microrheological measurement of the shear modulus}. 


In summary, we considered rigid spherical particles displaced against a surrounding elastic matrix by externally induced forces. 
We derived analytical expressions to calculate the resulting particle displacements. 
Mutual interactions 
due to induced matrix deformations are effectively included. 
This renders the procedure a promising tool 
to describe the behavior of elastic composite materials \cite{ilg2013stimuli,*odenbach2016microstructure,*menzel2016hydrodynamic}. 
Our experiments on 
paramagnetic particles 
in a soft elastic gel matrix and subject to tunable magnetic interactions confirm the potential of the theory.


Upon dynamic extension, a prospective application concerns macroscopic rheology 
\cite{denn2014rheology}, or nano- and microrheology 
\cite{ziemann1994local,*bausch1999measurement,
*waigh2005microrheology,*wilhelm2008out} 
where 
the 
matrix properties are tested by external agitation of embedded probe particles. 
Also biological and medical questions are addressable in this way, for instance cytoskeletal properties 
\cite{ziemann1994local,*bausch1999measurement,
*waigh2005microrheology,*wilhelm2008out,mizuno2007nonequilibrium}. 
An extension of the theory to include imposed torques \rev{on the particles, e.g., due to magnetic anisotropy,} is straightforward and will be presented in the near future.

\begin{acknowledgments}
The authors thank J.~Nowak for measuring the magnetization curves and the Deutsche Forschungsgemeinschaft for support of this work through the priority program SPP 1681 (Nos. AU 321/3-2 and ME 3571/3-2). 
\end{acknowledgments}


%
%
%

\begin{widetext}

\section*{
Supplemental material 
}

%
%
%


{\small
As stressed in the main text, the derivation of the displaceability matrix can likewise be performed for \textit{compressible} systems. Following the same steps of derivation as in the main text, we present below the corresponding expressions for completeness. Apart from that, we add further experimental results and comparison with the theory for a four-particle system, in complete analogy to our presentation for the three-particle system in the main text.
}

\vspace{1.cm}

\end{widetext}

\maketitle

\subsection{Expressions for a compressible elastic matrix}

For clarity and to facilitate the comparison with the hydrodynamic fluid case, we have presented in the main text the expressions for an \textit{incompressible} elastic system. That is, the system tends to locally preserve the volume of all its volume elements during any type of elastic deformation. However, and in contrast to the hydrodynamic fluid case \cite{batchelor1972hydrodynamic,karrila1991microhydrodynamics,dhont1996introduction}, for elastic matrices it is straightforward to allow for \textit{compressibility} in the derivation. This extended derivation proceeds in direct analogy to the one presented in the main text. 

We again assume a homogeneous and isotropic elastic matrix of infinite extension. Once more, we start from the linear elastostatic Navier-Cauchy equations \cite{cauchy1828exercises}, 
\begin{equation}\label{suppl_eq_navier-cauchy}
	\nabla^2\mathbf{u}(\mathbf{r}) + \frac{1}{1-2\nu}\nabla\nabla\cdot\mathbf{u}(\mathbf{r}) ={} -\frac{1}{\mu}\mathbf{f}_b(\mathbf{r}). 
\end{equation}
As in the main text, $\mathbf{u}(\mathbf{r})$ denotes the displacement field, 
$\mu$ the shear modulus of the matrix \cite{landau1986theory}, $\nu$ the Poisson ratio \cite{landau1986theory}, and $\mathbf{f}_b(\mathbf{r})$ the bulk force density. Now, 
we do not restrict our analysis to incompressible materials that locally adhere to $\nabla\cdot\mathbf{u}(\mathbf{r})=0$, and we do not assign a specific value to $\nu$. 

The resulting Green's function for a point force density $\mathbf{f}_b(\mathbf{r})=\mathbf{F}\delta(\mathbf{r})$ then reads \cite{landau1986theory}
\begin{equation}\label{suppl_greens_function}
	\mathbf{\hspace{.02cm}\underline{\hspace{-.02cm}G}}(\mathbf{r}) ={} \frac{1}{16\pi(1-\nu)\mu}\left[\frac{3-4\nu}{r}\mathbf{\underline{\hat{I}}}+\frac{\mathbf{r}\mathbf{r}}{r^3}\right].
\end{equation}
Using this expression, if an external force $\mathbf{F}_i$ is acting on a rigid spherical particle $i$ of radius $a$ embedded in the matrix with no-slip boundary conditions on its surface, a displacement field 
\begin{equation}\label{suppl_eq_u_i_0}
	\mathbf{u}_i^{(0)} (\mathbf{r}) = \left(1+\frac{a^2}{6}\nabla^2\right)\mathbf{\hspace{.02cm}\underline{\hspace{-.02cm}G}}(\mathbf{r}-\mathbf{r}_i)\cdot\mathbf{F}_i
\end{equation}
is induced. Eq.~(\ref{suppl_eq_u_i_0}) has the same form as in the main text, but $\mathbf{\hspace{.02cm}\underline{\hspace{-.02cm}G}}$ is different, see Eq.~(\ref{suppl_greens_function}). 
Again, the validity of Eq.~(\ref{suppl_eq_u_i_0}) is confirmed 
as it satisfies Eq.~(\ref{suppl_eq_navier-cauchy}), shows the correct limit $\mathbf{u}_i^{(0)}(\mathbf{r})=\mathbf{\hspace{.02cm}\underline{\hspace{-.02cm}G}}(\mathbf{r}\rev{-\mathbf{r}_i})\cdot\mathbf{F}_i$ for \rev{$|\mathbf{r}-\mathbf{r}_i|>a$ when} $a\rightarrow0$, and for $\rev{|\mathbf{r}-\mathbf{r}_i|}=a$ is constant on the particle surface. 
For $\rev{|\mathbf{r}-\mathbf{r}_i|}=a$, it reveals the rigid displacement 
\begin{equation}
\mathbf{U}_i^{(0)} \,=\, \mathbf{u}_i^{(0)}(\rev{|\mathbf{r}-\mathbf{r}_i|}=a) \,=\, \frac{5-6\nu}{24\pi(1-\nu)\mu a}\mathbf{F}_i
\end{equation} 
of the $i$th particle in response to $\mathbf{F}_i$. 

The no-slip condition under our assumptions applies on the surface $\partial V_j$ of a particle $j$ also for compressible matrices. Thus Eq.~(5) in the main text preserves its shape, i.e.\ 
\begin{equation}\label{suppl_eq_balance}
\mathbf{U}_j^{(1)}\!+\mathbf{\Omega}_j^{(1)}\!\times(\mathbf{r}-\mathbf{r}_j)=\mathbf{u}_i^{(0)}(\mathbf{r})+ \!\int_{\partial V_j}\!\mathbf{\hspace{.02cm}\underline{\hspace{-.02cm}G}}(\mathbf{r}-\mathbf{r}')\cdot\mathbf{f}(\mathbf{r}')\mathrm{d}S',
\end{equation}
where $\mathbf{U}_j^{(1)}$ denotes the translation of particle $j$, $\mathbf{\Omega}_j^{(1)}$ is its rotation, the displacement field $\mathbf{u}_i^{(0)}(\mathbf{r})$ is induced by particle $i$, and $\mathbf{f}(\mathbf{r}')$ denotes the surface force density that particle $j$ exerts on the surrounding matrix. The derivation of expressions for $\mathbf{U}_j^{(1)}$ and $\mathbf{\Omega}_j^{(1)}$ in the form of the Fax\'en laws follows the same strategy as described in the main text and leads to 
\begin{eqnarray}
		\mathbf{U}_j^{(1)} &={} &\left(1+\frac{a^2}{6}\nabla^2\right)\mathbf{u}_i^{(0)}(\mathbf{r})\bigg|_{\mathbf{r}=\mathbf{r}_j},\label{suppl_eq_faxen-translation} \\
		\boldsymbol{\Omega}_j^{(1)} &={} &\frac{1}{2}\nabla\times\mathbf{u}_i^{(0)}(\mathbf{r})\bigg|_{\mathbf{r}=\mathbf{r}_j}\label{suppl_eq_faxen-rotation}.
\end{eqnarray}

Also the stresslet $\mathbf{\underline{S}}_j$ exerted by particle $j$ onto the matrix is derived in analogy to what is described in the main text. In general, for compressible systems, this stresslet is given by the expression $\mathbf{\underline{S}}_j=\int_{\partial V_j}\mathrm{d}S'[\mathbf{f}(\mathbf{r}')\mathbf{r}'+
(\mathbf{f}(\mathbf{r}')\mathbf{r}')^T]/2$. This expression slightly differs from the one introduced below Eq.~(7) in the main text for incompressible systems. There, a trace-free definition was used 
to exclude compressions and dilations of the matrix, 
which needs to be the case for volume-conserving systems. It can be seen 
from the main text that the difference in definitions plays no actual role for our derivation. The reason is Eq.~(9), where the extra term $\sim\mathbf{\underline{\hat{I}}}$ in the incompressible case only leads 
to a contribution $\sim\nabla\cdot\mathbf{\hspace{.02cm}\underline{\hspace{-.02cm}G}}$. Yet, $\nabla\cdot\mathbf{\hspace{.02cm}\underline{\hspace{-.02cm}G}}$ vanishes in the incompressible case. Therefore, following the same strategy as described in the main text, we obtain
\begin{eqnarray}
	\mathbf{\underline{S}}_j^{(1)} &\!=\!&{} -\frac{4\pi(1-\nu)\mu a^3}{4-5\nu}\left(1+\frac{a^2}{10}\nabla^2\right)
\!
\Bigg[\frac{1}{1-2\nu}\mathbf{\underline{\hat{I}}}\,\nabla\!\cdot\!\mathbf{u}_i^{(0)}(\mathbf{r})
\notag\\ 
&&{}+\frac{5}{2}\Big(\nabla\mathbf{u}_i^{(0)}(\mathbf{r})+\big(\nabla\mathbf{u}_i^{(0)}(\mathbf{r})\big)^T\Big) \Bigg]\Bigg|_{\mathbf{r}=\mathbf{r}_j}. \label{suppl_eq_faxen-stresslet}
\end{eqnarray}
Likewise, the displacement field $\mathbf{u}_j^{(1)}(\mathbf{r})$ resulting from the rigidity of particle $j$ and its resistance to deformation, expressed by the stresslet $\mathbf{\underline{S}}_j^{(1)}$, is calculated as described in the main text. Eq.~(\ref{suppl_eq_faxen-stresslet}) here contains a term $\sim 1/(1-2\nu)$, which would diverge for $\nu\rightarrow 0.5$. However, it gets canceled by a counter-factor $\sim(1-2\nu)$ in the calculation. More precisely, upon inserting Eq.~(\ref{suppl_eq_u_i_0}) into Eq.~(\ref{suppl_eq_faxen-stresslet}), the expression $\nabla\cdot\mathbf{\hspace{.02cm}\underline{\hspace{-.02cm}G}}$ appears; straightforward calculation of $\nabla\cdot\mathbf{\hspace{.02cm}\underline{\hspace{-.02cm}G}}$ via Eq.~(\ref{suppl_greens_function}) leads to a factor $\sim(1-2\nu)$. In the end, $\mathbf{u}_j^{(1)}(\mathbf{r})$ has the same form as Eq.~(9) in the main text, 
\begin{equation}\label{suppl_eq_u_j_1}
	\mathbf{u}_j^{(1)}(\mathbf{r}) ={} -\left(\mathbf{\underline{S}}_j^{(1)}\cdot\nabla\right)\cdot\mathbf{\hspace{.02cm}\underline{\hspace{-.02cm}G}}(\mathbf{r}-\mathbf{r}_j).
\end{equation}

In the next step, again, the reaction of particle $i$ in response to the field $\mathbf{u}_j^{(1)}(\mathbf{r})$ is obtained from Eqs.~(\ref{suppl_eq_faxen-translation})--(\ref{suppl_eq_u_j_1}) by replacing $(\mathbf{u}_i^{(0)}, \mathbf{U}_j^{(1)}, \mathbf{\Omega}_j^{(1)}, \mathbf{\underline{S}}_j^{(1)}, \mathbf{u}_j^{(1)}, \mathbf{r}_j)$ with $(\mathbf{u}_j^{(1)}, \mathbf{U}_i^{(2)}, \mathbf{\Omega}_i^{(2)}, \mathbf{\underline{S}}_i^{(2)}, \mathbf{u}_i^{(2)}, \mathbf{r}_i)$. 

Summing up the contributions from the different steps of iteration and considering all $N$ particles simultaneously leads to an expression in the form of an elastic displaceability matrix $\mathbf{\underline{M}}$ as given in the main text: 
%
%
%
\begin{eqnarray}\label{suppl_eq_displaceability_matrix}
\begin{pmatrix}
	\mathbf{U}_1\\
	\vdots\\
	\mathbf{U}_N
\end{pmatrix}
={}
\begin{pmatrix}
	\mathbf{\underline{M}}_{11} &   \ldots & \mathbf{\underline{M}}_{1N} \\
	\vdots & \vdots & \vdots  \\
	\mathbf{\underline{M}}_{N1} & \ldots & \mathbf{\underline{M}}_{NN}
\end{pmatrix}
\cdot
\begin{pmatrix}
	\mathbf{F}_1\\
	\vdots\\
	\mathbf{F}_N
\end{pmatrix}.
\end{eqnarray}
Limiting ourselves to contributions up to (including) order $r_{ij}^{-4}$, we find for a \textit{compressible} system the more general expressions
\begin{eqnarray}
	\mathbf{\underline{M}}_{i=j} &\!=\!& M_0\Bigg\{\mathbf{\underline{\hat{I}}} -  \sum\limits_{\substack{k=1 \\	k\not=i}}^{N} \frac{3}{4(4-5\nu)(5-6\nu)}\bigg(\frac{a}{r_{ik}}\bigg)^{\!\!\!4}\notag\\
	&{} &\Big[ \Big(37-44\nu+10(1-2\nu)^2\Big)\mathbf{\hat{r}}_{ik}\mathbf{\hat{r}}_{ik}\notag\\
		&&{}+5(1-2\nu)^2\left(\mathbf{\underline{\hat{I}}}- \mathbf{\hat{r}}_{ik}\mathbf{\hat{r}}_{ik}\right) \Big] \Bigg\}, \label{suppl_M_II_2}\\
	\mathbf{\underline{M}}_{i\not=j} &\!=\!& M_0 \frac{3}{2(5-6\nu)}\frac{a}{r_{ij}}  \Bigg[ \Bigg( 4(1-\nu)-\frac{4}{3} \bigg(\frac{a}{r_{ij}}\bigg)^{\!\!\!2} \Bigg) \mathbf{\hat{r}}_{ij}\mathbf{\hat{r}}_{ij}  \notag\\ 
			& & {}+ \Bigg( 3-4\nu+\frac{2}{3} \bigg(\frac{a}{r_{ij}} \bigg)^{\!\!\!2} \Bigg)\!\left(\mathbf{\underline{\hat{I}}}-\mathbf{\hat{r}}_{ij}\mathbf{\hat{r}}_{ij}\right)\! \Bigg]
+ \mathbf{\underline{M}}_{i\not=j}^{(3)}    ,
\label{suppl_M_IJ_2} 
\notag\\[-.2cm]&&
\end{eqnarray}
where $M_0 ={} (5-6\nu)/24\pi(1-\nu)\mu a$ and $\mathbf{\hat{r}}_{ij}=\mathbf{r}_{ij}/r_{ij}$ ($i,j=1,2,...,N$). 
Here, the three-body interactions contribute as given by $\mathbf{\underline{M}}_{i\not=j}^{(3)}$ in the form 
\begin{eqnarray}\label{suppl_M_tt_3}
	\mathbf{\underline{M}}_{i\not=j}^{(3)} &\!=\!& M_0\frac{3}{8(4-5\nu)(5-6\nu)}
	\sum\limits_{\substack{k=1\\k\not=i,j}}^{N}
	\bigg(\frac{a}{r_{ik}}\bigg)^{\!\!2}
	\bigg(\frac{a}{r_{jk}}\bigg)^{\!\!2}
\notag\\	
	&&\Big[
		\!-\!10(1-2\nu)\Big( (1-2\nu)\big((\mathbf{\hat{r}}_{ik}\cdot\mathbf{\hat{r}}_{jk}) \mathbf{\underline{\hat{I}}}+ \mathbf{\hat{r}}_{jk}\mathbf{\hat{r}}_{ik} \big) \notag\\
		 &{}&+ 3 (\mathbf{\hat{r}}_{ik}\cdot\mathbf{\hat{r}}_{jk}) (\mathbf{\hat{r}}_{ik}\mathbf{\hat{r}}_{ik} + \mathbf{\hat{r}}_{jk}\mathbf{\hat{r}}_{jk}) - \mathbf{\hat{r}}_{ik}\mathbf{\hat{r}}_{jk} \Big) \notag\\
			&{}&+ 3\left(7-4\nu - 15 (\mathbf{\hat{r}}_{ik}\cdot\mathbf{\hat{r}}_{jk})^2\right)\mathbf{\hat{r}}_{ik}\mathbf{\hat{r}}_{jk}
		\Big]. 
\end{eqnarray}

The corresponding expressions for \textit{incompressible} systems in the main text readily follow from Eqs.~(\ref{suppl_eq_displaceability_matrix})--(\ref{suppl_M_tt_3}) by setting the Poisson ratio $\nu=0.5$. Here, we derived and listed the more general expressions for \textit{compressible} elastic matrices.



\subsection{Four-particle system}


In addition to the two- and three-particle samples described in the main text, we also generated and analyzed four-particle systems. Their preparation, experimental analysis, and the corresponding comparison with the theory are in complete analogy to the three-particle system described in the main text. 
We recall that our theoretical description in the main text \textit{up to the investigated order} (including $r_{ij}^{-4}$) is exact for arbitrary particle numbers. No higher-body interactions appear to this order. Therefore, Eqs.~(10)--(14) in the main text also apply to systems of particle numbers $N>3$ up to (including) order $r_{ij}^{-4}$, i.e.\ if the particle separations are not significantly reduced. 

\begin{figure*}[t!]
\includegraphics[width=.8\textwidth]{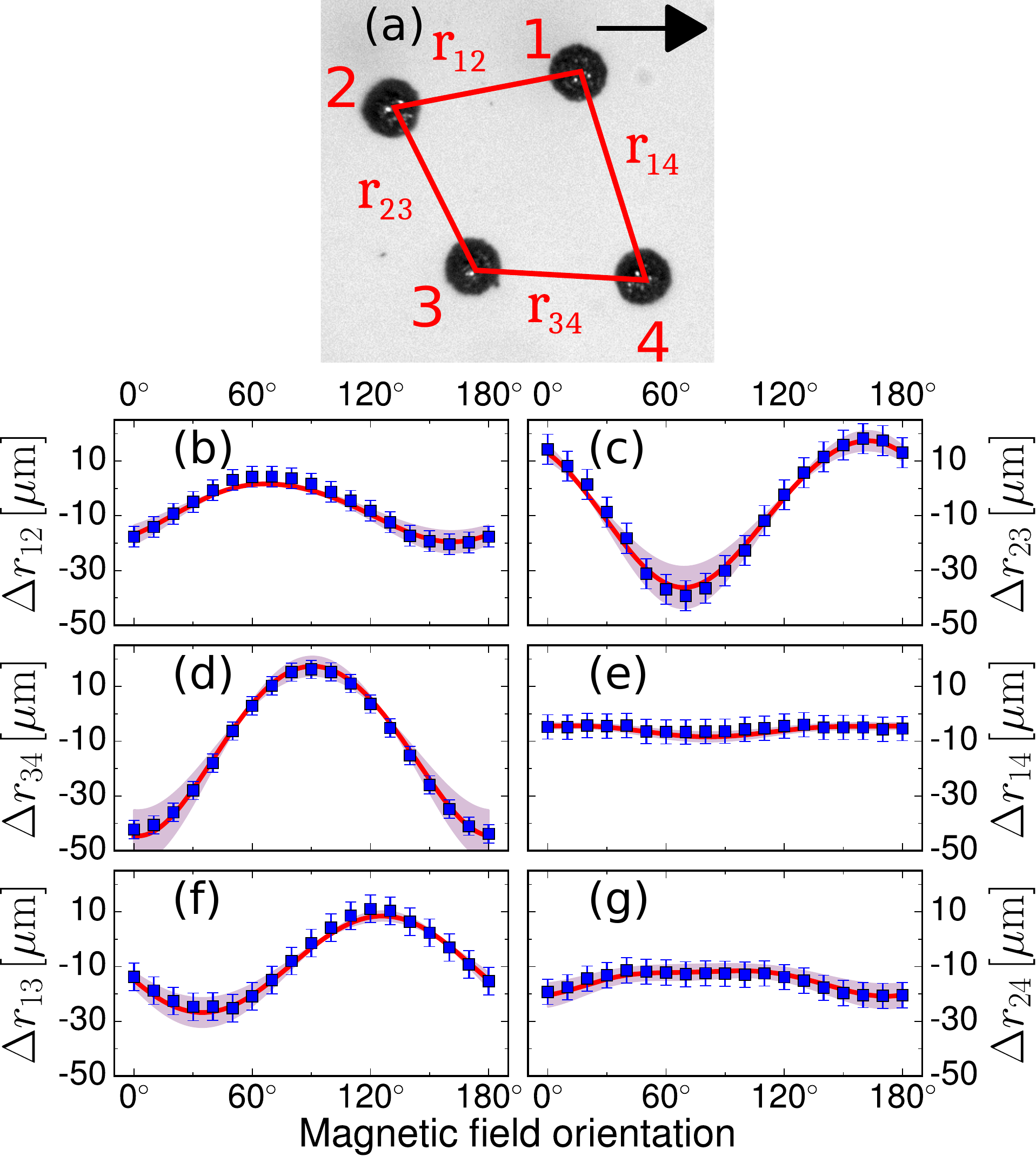}
\caption{Same as in Figs.~3 and 4 in the main text, here for a four-particle system. (a) The snapshot was taken for vanishing external magnetic field (particle diameters {204.4}$\pm${2.2}~$\mathrm{\mu m}$). (b--g) Changes $\Delta r_{ij}$ in all six distances ($i,j=1,2,3,4$, $i\neq j$). 
Good agreement between theory (red line) and experiments (blue squares) is observed, and
the modulus of the gel matrix for this system is obtained as {85.7}$\pm${12.6}~Pa.}
\label{fig_4particle}
\end{figure*}

Thus, our four-particle results predominantly provide a supplement to the results presented in the main text. 
Our experimental and theoretical results for the four-particle system are depicted in Fig.~\ref{fig_4particle}. 
One could continue to further increasing particle numbers in the same way. 

\newpage


\end{document}